\input harvmac

\def\Title#1#2{\rightline{#1}\ifx\answ\bigans\nopagenumbers\pageno0\vskip1in
\else\pageno1\vskip.8in\fi \centerline{\titlefont #2}\vskip .5in}
%

%
%
\ifx\includegraphics\UnDeFiNeD\message{(NO graphicx.tex, FIGURES WILL BE IGNORED)}
\def\figin#1{\vskip2in}
\else\message{(FIGURES WILL BE INCLUDED)}\def\figin#1{#1}
\fi
\def\Fig#1{Fig.~\the\figno\xdef#1{Fig.~\the\figno}\global\advance\figno
 by1}
%
%
%
%

%
%
\font\ticp=cmcsc10

\def\calo{{\cal O}}
\def\calh{{\cal H}}

\def\hcore{\calh_{\rm core}}
\def\hreg{\calh_{\rm reg}}
\def\hnear{\calh_{\rm near}}
\def\hfar{\calh_{far}}

\def\roughly#1{\mathrel{\raise.3ex\hbox{$#1$\kern-.75em\lower1ex\hbox{$\sim$}}}}

\def\ala{{\overleftarrow a}}
\def\ara{{\overrightarrow a}}
\def\Ula{{\overleftarrow U}}
\def\Ura{{\overrightarrow U}}
\def\ula{{\overleftarrow u}}
\def\ura{{\overrightarrow u}}
\def\nla{{\overleftarrow n}}
\def\nra{{\overrightarrow n}}
\overfullrule=0pt
%
%

\lref\Hawkrad{
  S.~W.~Hawking,
  ``Particle Creation By Black Holes,''
  Commun.\ Math.\ Phys.\  {\bf 43}, 199 (1975)
  [Erratum-ibid.\  {\bf 46}, 206 (1976)].
}
\lref\BHMR{
  S.~B.~Giddings,
  ``Black holes and massive remnants,''
Phys.\ Rev.\  {\bf D46}, 1347-1352 (1992).
[hep-th/9203059].
}
\lref\thooholo{
  G.~'t Hooft,
  ``Dimensional reduction in quantum gravity,''
  arXiv:gr-qc/9310026.
}
\lref\sussholo{
  L.~Susskind,
  ``The World As A Hologram,''
  J.\ Math.\ Phys.\  {\bf 36}, 6377 (1995)
  [arXiv:hep-th/9409089].
}
\lref\AMPS{
  A.~Almheiri, D.~Marolf, J.~Polchinski and J.~Sully,
  ``Black Holes: Complementarity or Firewalls?,''
[arXiv:1207.3123 [hep-th]].
}
\lref\HaHa{
  D.~Harlow and P.~Hayden,
  ``Quantum Computation vs. Firewalls,''
[arXiv:1301.4504 [hep-th]].
}
\lref\SusskindTG{
  L.~Susskind,
  ``Black Hole Complementarity and the Harlow-Hayden Conjecture,''
[arXiv:1301.4505 [hep-th]].
}
\lref\GiddingsSJ{
  S.~B.~Giddings,
  ``Black hole information, unitarity, and nonlocality,''
Phys.\ Rev.\ D {\bf 74}, 106005 (2006).
[hep-th/0605196].
}
\lref\SBGmodels{
  S.~B.~Giddings,
  ``Models for unitary black hole disintegration,''
Phys.\ Rev.\ D {\bf 85}, 044038 (2012).
[arXiv:1108.2015 [hep-th]].
}
\lref\PageUP{
  D.~N.~Page,
  ``Black hole information,''
[hep-th/9305040].
}
\lref\GiddingsUE{
  S.~B.~Giddings,
  ``Black holes, quantum information, and unitary evolution,''
Phys.\ Rev.\ D {\bf 85}, 124063 (2012).
[arXiv:1201.1037 [hep-th]].
}
\lref\Mathur{
  S.~D.~Mathur,
  ``The Information paradox: A Pedagogical introduction,''
Class.\ Quant.\ Grav.\  {\bf 26}, 224001 (2009).
[arXiv:0909.1038 [hep-th]]; ``What the information paradox is {\it not},''
[arXiv:1108.0302 [hep-th]].
}
\lref\Compl{
  L.~Susskind, L.~Thorlacius and J.~Uglum,
  ``The Stretched horizon and black hole complementarity,''
Phys.\ Rev.\ D {\bf 48}, 3743 (1993).
[hep-th/9306069].
}
\lref\BaFi{
  T.~Banks,
  ``Holographic Space-Time: The Takeaway,''
[arXiv:1109.2435 [hep-th]].
}
\lref\HaydenCS{
  P.~Hayden, J.~Preskill and ,
  ``Black holes as mirrors: Quantum information in random subsystems,''
JHEP {\bf 0709}, 120 (2007).
[arXiv:0708.4025 [hep-th]].
}
\lref\GiddingsDH{
  S.~B.~Giddings, Y.~Shi and ,
  ``Quantum information transfer and models for black hole mechanics,''
[arXiv:1205.4732 [hep-th]].
}
\lref\SusskindUW{
  L.~Susskind,
  ``The Transfer of Entanglement: The Case for Firewalls,''
[arXiv:1210.2098 [hep-th]].
}
\lref\LaMa{
  A.~E.~Lawrence and E.~J.~Martinec,
  ``Black hole evaporation along macroscopic strings,''
Phys.\ Rev.\ D {\bf 50}, 2680 (1994).
[hep-th/9312127].
}
\lref\Brown{
  A.~R.~Brown,
 ``Tensile Strength and the Mining of Black Holes,''
[arXiv:1207.3342 [gr-qc]].
}
\lref\algquant{
  R.~Haag,
  {\sl Local quantum physics: Fields, particles, algebras,}
Berlin, Germany: Springer (1992) 356 p. (Texts and monographs in physics).
}
\lref\dewitt{
  B.~S.~DeWitt,
  ``Quantum Field Theory in Curved Space-Time,''
Phys.\ Rept.\  {\bf 19}, 295 (1975).
}
\lref\GiNe{
  S.~B.~Giddings and W.~M.~Nelson,
 ``Quantum emission from two-dimensional black holes,''
Phys.\ Rev.\ D {\bf 46}, 2486 (1992).
[hep-th/9204072].
}
\lref\Page{
  D.~N.~Page,
  ``Average entropy of a subsystem,''
Phys.\ Rev.\ Lett.\  {\bf 71}, 1291 (1993).
[gr-qc/9305007];
 ``Information in black hole radiation,''
  Phys.\ Rev.\ Lett.\  {\bf 71}, 3743 (1993)
  [arXiv:hep-th/9306083].
}
\lref\GiShWIP{S.B.~Giddings and Y.~Shi, work in progress.}
\lref\UnWa{
  W.~G.~Unruh and R.~M.~Wald,
  ``Acceleration Radiation and Generalized Second Law of Thermodynamics,''
Phys.\ Rev.\ D {\bf 25}, 942 (1982);
``How to mine energy from a black hole," Gen. Rel. and Grav. {\bf 15}, 195 (1983).
}
\lref\Frol{
  V.~P.~Frolov and D.~Fursaev,
  ``Mining energy from a black hole by strings,''
Phys.\ Rev.\ D {\bf 63}, 124010 (2001).
[hep-th/0012260]\semi
  V.~P.~Frolov,
 ``Cosmic strings and energy mining from black holes,''
Int.\ J.\ Mod.\ Phys.\ A {\bf 17}, 2673 (2002).
}
\lref\NVNL{
  S.~B.~Giddings,
  ``Nonviolent nonlocality,''
[arXiv:1211.7070 [hep-th]].
}
\lref\KlebanovEH{
  I.~R.~Klebanov, L.~Susskind and T.~Banks,
  ``Wormholes And The Cosmological Constant,''
Nucl.\ Phys.\ B {\bf 317}, 665 (1989)..
}
\lref\Cole{
  S.~R.~Coleman,
``Black Holes as Red Herrings: Topological Fluctuations and the Loss of Quantum Coherence,''
Nucl.\ Phys.\ B {\bf 307}, 867 (1988).
}
\lref\GiSt{
  S.~B.~Giddings and A.~Strominger,
  ``Loss of Incoherence and Determination of Coupling Constants in Quantum Gravity,''
Nucl.\ Phys.\ B {\bf 307}, 854 (1988).
}
\lref\NLvsC{
  S.~B.~Giddings,
  ``Nonlocality versus complementarity: A Conservative approach to the information problem,''
Class.\ Quant.\ Grav.\  {\bf 28}, 025002 (2011).
[arXiv:0911.3395 [hep-th]].
}
\Title{
\vbox{\baselineskip12pt
}}
{\vbox{
\centerline{Nonviolent information transfer from black holes:}\centerline {a field theory parameterization}
}}
\centerline{{\ticp Steven B. Giddings\footnote{$^\ast$}{Email address: giddings@physics.ucsb.edu}  }
}
\centerline{\sl Department of Physics}
\centerline{\sl University of California}
\centerline{\sl Santa Barbara, CA 93106}
\vskip.40in
\centerline{\bf Abstract}
 A candidate parameterization is introduced, in an effective field theory framework, for the quantum information transfer from a black hole that is necessary to restore unitarity.  This in particular allows description of the effects of this information transfer in the black hole atmosphere, for example seen by infalling observers.  In the presence of such information transfer, it is shown that infalling observers need not experience untoward violence.  Moreover, the presence of general moderate-frequency couplings to field modes with high angular momenta offers a mechanism to enhance information transfer rates, commensurate with the increased energy flux, when a string is introduced to ``mine" a black hole.  Generic such models for nonviolent information transfer predict extra energy flux from a black hole, beyond that of Hawking.

\vskip.3in
\Date{}

In the semiclassical geometric picture, quantum information that falls into a black hole (BH) never returns.  Likewise, Hawking evaporation\Hawkrad\ can be described as creation of correlated particles near the horizon, analogous to an EPR pair, with one member of the pair escaping and the other lost to the inside of the black hole.  In either case the inside information can never be recovered from the black hole by the locality principle of local quantum field theory(LQFT):  quantum information does not propagate faster than the speed of light.  This has provoked an apparent crisis in physics.

If locality forbids quantum information transfer from a black hole, we must go beyond the LQFT framework to describe the quantum information transfer necessary to save quantum mechanics.  One idea for this is formation of an interface that propagates from the deep interior to the horizon of a large black hole, transferring quantum information and producing a ``star-like" massive remnant\BHMR.  Another is the more radical  
idea that perhaps the information was never that localized to begin with, {\it e.g.} due to a new form of complementarity\refs{\thooholo\sussholo-\Compl}.  

A crucial question is the nature of the information transfer or other nonlocality, as departure from LQFT.  One possibility is that such violation stops right {\it at} the horizon of a black hole, or a Planck distance away.  This assumption is postulate II of BH complementarity\Compl.  As we will review, this postulate results in dire consequences, if the missing information is encoded in outgoing radiation.  This problem, which is easily seen by tracing outgoing perturbations of the Hawking state back to the horizon (see, {\it e.g.}, \refs{\GiddingsSJ\Mathur\SBGmodels-\GiddingsUE}),  has been refined and is now commonly known as the ``firewall" problem\AMPS.

Nonlocality is a basic feature of the scenario described in \refs{\BHMR,\AMPS}:  if we begin with a black hole, information must transfer from its deep interior to its horizon.  However, with postulate II, the nonlocality stops exactly there.  If we consider the biggest black holes known in nature, in the firewall scenario information must be able to nonlocally propagate ten times the distance to Neptune -- but not a Planck length further, for a black hole of that radius.\foot{In contrast, more recent proposals to save a version of complementarity\refs{\HaHa,\SusskindTG} may involve a breakdown of locality to beyond the Andromeda galaxy, for a solar mass black hole located here.}  

Instead, \refs{\SBGmodels,\GiddingsUE,\NVNL} proposed that information transfer extends also to the immediate atmosphere of the black hole, dropping postulate II.  This was proposed to avoid the potential problem of singular horizons, so might be referred to as ``nonviolent nonlocality."  

The central question is the fundamental framework which replaces LQFT and explain such behavior.  It seems plausible that this is not based on a fundamental spacetime description, but does have a coarser notion of locality, {\it e.g.} encoded in evolution in a Hilbert space (or nested Hilbert spaces) with a certain kind of subsystem structure {\it e.g.} mimicking a manifold.  Such a picture may extend ideas of algebraic quantum field theory\algquant.  A proposal in this spirit is Banks and Fischler's holographic spacetime\BaFi; another with important differences is outlined in \refs{\GiddingsUE}.  While formulating the ultimate laws of quantum gravity is beyond the scope of this paper, a plausible view is that the unitarity crisis for black holes (and related problems in cosmology) may serve as a critical {\it guide}, just as the crisis of atomic stability did for quantum mechanics.  

This paper instead takes the viewpoint that while profound departure from LQFT may be necessary, the ultimate consequences of that departure may appear as a small departure from LQFT, in an appropriate sense, in regions such as a black hole atmosphere.  Thus, following \NVNL, it will discuss parameterization of departures from locality, in the LQFT framework.  

To give such a description, first consider the usual semiclassical LQFT evolution of a nonrotating black hole\foot{For present discussion, we neglect time evolution, since ${\delta R/ R} \sim 1/R^2$, due to Hawking emission, over time $\sim R$, where $R$ is the BH radius.},   
\eqn\BHmet{ds^2 = -f(r) dt^2 + {dr^2\over f(r)} + r^2 d\Omega^2\ ,}
where for four-dimensional Schwarzschild,
\eqn\fdf{f(r) = 1-{R\over r}\ .}
The evolution can be described by introducing a time slicing; let's take time slices to match $t$ at infinity, but smoothly cut across the horizon and hit the classical singularity.  Such a slicing has been called\NLvsC\  ``natural," since it arises from a collection of  clocks, starting at different radii, that fall into the black hole.  On a given time slice, away from the singularity, small perturbations are well-described by LQFT.  Moreover, locality implies a decomposition into different subsystems corresponding to different regions on the slice. 
A coarse version of the decomposition\refs{\SBGmodels,\GiddingsUE} is into the interior system, with $r<R$, the atmosphere, with $R<r\roughly<2R$, and the far exterior, with $2R\roughly< r$.  Bearing in mind the manifest breakdown of LQFT at $r=0$, we might further divide the interior into a ``core" system at $r<R_c$, for some $R_c\ll R$, and the complementary regular region $R_c<r<R$.  A corresponding Hilbert space decomposition is of the form $\hcore\otimes\hreg\otimes\hnear\otimes\hfar$.  Evolution via the LQFT hamiltonian propagates excitations between the regions, except outwards from $\hcore$ and $\hreg$.  

A simple hypothesis, in order to save unitarity, is that the true physics introduces some critical modifications.   Perhaps the most important is a term allowing information transfer\foot{This can be characterized in terms of transfer of entanglement\refs{\HaydenCS,\GiddingsDH,\SusskindUW}.} from $\hcore$ to $\hreg\otimes\hnear$, which we will parameterize with a hamiltonian $H_{trans}$.  One also expects states and evolution in $\hcore$ to have significant departures from LQFT evolution.\foot{Ref.~\NVNL\ suggested a LQFT parameterization of core states in terms of states on a ``nice slice" asymptotic to radius $<R_c$.}  The latter could be represented by a hamiltonian $H_{sc}$, scrambling core states at some rate.  One may also consider (or be required to consider) other modifications to LQFT evolution, {\it e.g.} transfer from $\hreg$ to $\hnear$, but we will focus on that due to $H_{trans}$.  

Ref.~\NVNL\ described such evolution in terms of creation/annihilation operators in a basis of wavepackets, {\it e.g.}
\eqn\waveH{ H_{trans} \sim {1\over R} a^\dagger_{near}{\cal N} a_{core} + h.c.\quad ,\quad  H_{sc} \sim {1\over R} a^\dagger_{core}{\cal S} a_{core}+ h.c.\ ,}
parameterized by matrices ${\cal N}$ and $\cal S$.  In field theory, such expressions may be rewritten in terms of operators acting on the appropriate Hilbert spaces.  Taking the case of general operators ${\cal O}_A$, we thus consider a hamiltonian of the form
\eqn\genH{ \int dt H_{NL} = \int dV_4 dV_4'{\cal O}_A(x) G_{AB}(x,x')  {\cal O}_B(x')}
or the action equivalent, parameterized by a matrix $G_{AB}(x,x')$ ($dV_4$ is the volume element).  The specific functional form of $G_{AB}$, which is taken to depend on the BH background, can in particular be used to parameterize information transfer from $\hcore$ to $\hnear$, for example equivalent to  $H_{trans}$  in \waveH, or with some other characteristic time scale.\foot{Note that if the full gravitational dynamics admits gauge transformations analogous to those in classical gravity, there could also be a gauge equivalent description corresponding to a Schwarzschild time slicing.  In such a description, which might realize the ``outside" picture common to discussions of complementarity, information would be relayed from the horizon to the atmosphere\NVNL.}    Since we expect LQFT evolution to ultimately fail badly for states in $\hcore$, one may alternately need a different basis of operators acting on that part of the system.   

Note that\NVNL\ the special case of constant $G_{AB}$'s corresponds to the type of interaction induced by spacetime wormholes;\foot{Indeed, Page has speculated that wormholes could transfer information between core and atmosphere regions\PageUP.} in that case, information is not transferred\refs{\Cole\GiSt-\KlebanovEH}.  But, spacetime-dependent $G_{AB}$ can clearly propagate information.  Another possible generalization is to allow ``interaction" terms that appear multilocal in terms of field theory operators.

The focus of the rest of this paper will be the effects of such nonlocal evolution on the state outside the BH.  If we work in an interaction picture where the interaction \genH\ transferring information from $\hcore$ to $\hnear$ is treated as the perturbation,\foot{In particular, one might also treat $H_{sc}$ as part of the {\it unperturbed} hamiltonian, to describe evolution of core states.} the state of the system evolves as
\eqn\evol{|\psi(t)\rangle = T\exp\left\{ -i \int dt H_{NL}(t)\right\}|\psi(0)\rangle\ .}
Specifically, consider expectation values of observables $\cal O$ outside the black hole, $\langle\psi(t)|{\cal O}|\psi(t)\rangle$.  These observables are only sensitive to the effects of the operators in $H_{NL}$ that act on the degrees of freedom outside the BH.  So, from the viewpoint of outside evolution, $H_{NL}$ is a hamiltonian with a local operator coupled to a source $J_A$ which depends on the internal state of the black hole, 
\eqn\effsource{  \int dV_4 dV_4'{\cal O}_A(x) G_{AB}(x,x')  {\cal O}_B(x') \rightarrow \int dV_4 J_A(x) {\cal O}_A(x)\ .}

We would like to understand the effects of such sources, an in particular the question of whether such a source,\foot{Note that, strictly speaking, the internal/external couplings of \effsource\ are not equivalent to a {\it classical} source, although that is not critical in the present discussion.} if it creates enough excitation to transfer the necessary ``missing" information, can avoid the problem of a singular horizon.  A simplest example is a linear coupling,
\eqn\phicoup{\int dV_4 J^I(x) \Phi^I(x)}
to each of the fields in nature.  Of course such an expression may be too na\"\i ve for standard model fields, {\it e.g.} due to the need to conserve gauge charges, {\it etc.}, but we will explore such couplings as a model for more general couplings.   Another example of a universal coupling to all fields is
\eqn\Tcoup{\int dV_4 J^{\mu\nu}(x) T_{\mu\nu}(x)\ ,}
where $T_{\mu\nu}$ is the stress tensor for the fields in nature.  This coupling would arise from metric fluctuations near the black hole ({\it e.g.} perhaps due to some  horizon fluctuation) if a mechanism existed to ``present" the missing quantum information in these.  Study of the effects of such a coupling will be deferred to future work, but is expected to have important similarities with the results from a linear coupling, \phicoup, to be described below.

To understand the effects of such sources, particularly in the black hole atmosphere, let us focus on the example of a single scalar field $\phi$ linearly coupled to a source $J(x)$,
\eqn\lincoup{\int dt H_{NL}\rightarrow \int dV_4 J(x) \phi(x), }
and investigate properties of the resulting state.

The effect of the source is easily described by expanding the field in a basis of modes.  These are efficiently characterized by passing to tortoise coordinates,
\eqn\tortdef{r^* = \int {dr\over f(r)}\ ;}
for four-dimensional Schwarzschild, \fdf, the tortoise coordinate $r^*$ is given by 
\eqn\tort{e^{{r^*/R}} = \left({r\over R} -1\right) e^{{r/ R}-1}\ .}
Then, scalar field modes with definite angular momentum, 
\eqn\sepl{\phi = Y_{lm}(\Omega) {u_l(r,t)\over r}}
satisfy a two-dimensional wave equation, with an effective potential:
\eqn\tdeqn{\left(-{\partial^2\over \partial t^2} + {\partial^2\over \partial r^{*2}} \right)u_l = V_l(r^*) u_l\ ,}
\eqn\effpot{V_l = \left(1-{R\over r}\right) \left[{l(l+1)\over r^2} + {R\over r^3}\right]\ .}
As $r^*\rightarrow \pm\infty$, $V_l\rightarrow0$ and solutions are $(1+1)$-dimensional plane waves.  Orthonormal modes are given by (see, {\it e.g.} \refs{\dewitt} )
\eqn\pwmodes{{\overrightarrow u}_{l,\omega}(t,r^*) = {1\over \sqrt{2\omega}} e^{-i\omega t} {\overrightarrow u}_{l,\omega}(r^*)\ ;\ {\overleftarrow u}_{l,\omega}(t,r^*) = {1\over \sqrt{2\omega}} e^{-i\omega t} {\overleftarrow u}_{l,\omega}(r^*)}
where 
\eqn\rmdef{\ura_{l,\omega}(r^*)\  {\buildrel r^*\rightarrow -\infty \over \longrightarrow}\  e^{i\omega r^*} + {\overrightarrow A}_{l\omega} e^{-i\omega r^*}\quad ;\quad {\buildrel r^*\rightarrow \infty \over \longrightarrow}\  B_{l\omega} e^{i\omega r^*}\ .}
and
\eqn\lmdef{\ula_{l,\omega}(r^*)\  {\buildrel r^*\rightarrow -\infty \over \longrightarrow}\  B_{l\omega} e^{-i\omega r^*}\  \quad ;\quad {\buildrel r^*\rightarrow \infty \over \longrightarrow}\  e^{-i\omega r^*} + {\overleftarrow A}_{l\omega} e^{i\omega r^*}\  .}
These may then be converted into a wavepacket basis, {\it e.g.} by 
\eqn\wavepdef{ \ura_{jn}(t,r^*) = {1\over \sqrt\epsilon} \int_{j\epsilon}^{(j+1)\epsilon} {d\omega}  e^{-i\omega(t+2\pi n/\epsilon)} \ura_{l,\omega}(r^*)\ ,}
and analogously for $\ula$, with integer $j$ and $n$, and $\epsilon$ an arbitrarily-chosen resolution parameter.
Then, a complete set of wavepacket modes  ${\overrightarrow U}_A$, ${\overleftarrow U}_A$, with collective index $A=\{jnlm\}$, is defined via \sepl, with Klein-Gordon norms
\eqn\KGnorms{ ( {\overrightarrow U}_A,{\overrightarrow U}_{A'})= \delta_{AA'} =({\overleftarrow U}_A,{\overleftarrow U}_{A'}), }
together with complex conjugate expressions, and other inner products vanishing.

The scalar field may be expanded in terms of such a basis, as
\eqn\phiexp{\phi(x) =\sum_A \left(  \ara_A \Ura_A+ \ala_A \Ula_A  \right) + h.c.\ ,}
and the state of the system may be characterized in terms of the corresponding occupation numbers.  Denote by $|0\rangle$ the Hawking (or Unruh) vacuum, and by $|\{\nra_A\},\{\nla_A\}\rangle$ the unit-normalized occupation number states built on this by acting with the $\ara^\dagger_A$'s and $\ala^\dagger_A$'s.

Important questions are what states are created by the interaction \lincoup, what states are sufficient to carry the needed black hole information, and how observably these states depart from the vacuum, particularly as seen by infalling observers.  
First consider the last question.  In the state $|\{\nra_A\},\{\nla_A\}\rangle$, the expectation value of the stress tensor is
\eqn\Tvev{\langle\{\nra_A\},\{\nla_A\}| T_{\mu\nu} |\{\nra_A\},\{\nla_A\}\rangle = \sum_A{ \nra_A t_{\mu\nu}( \Ura_A) + \nla_A t_{\mu\nu}(  \Ula_A)\over 1-e^{- 4\pi R\omega_A}} + T_{0\mu\nu}\ ,}
where $t_{\mu\nu}(f) = \partial_\mu f^*\partial_\nu f - g_{\mu\nu} \partial f^*\cdot \partial f/2$, and $T_{0\mu\nu}$ is the stress tensor of the Hawking radiation.  

For an outgoing mode with energy $\omega_A=\omega_j \simeq j\epsilon$, and width $\delta r^*\simeq 1/\epsilon$, the stress tensor describes an energy density $\sim \nra_{jnlm}|B_{l\omega_j}|^2\omega_j/(4\pi r^2 \delta r^*)$ propagating to infinity.  The gray body factor $B_{l\omega}$ is of order unity for $\omega R\roughly> l$, and typical Hawking modes have $\omega\sim 1/R$.

If missing information is to be carried by such modes, a benchmark rate, in order to restore missing correlations to the exterior state, is one extra quantum emitted per time $\sim R$.  If the information-carrying quanta also have frequency $\sim 1/R$ the power radiated is comparable to the Hawking radiation, which is a small effect for a big black hole; even radiation of quanta with $\omega \sim  1/R^p$ give parametrically small energy densities. One also has significant latitude for quanta to be emitted more rapidly.

Another possible benchmark is the needed information emission rate to match the maximal rate at which a black hole can be mined, by a collection of cosmic strings, as described in \refs{\UnWa\LaMa-\Frol,\AMPS,\Brown}.  According to \Brown, the maximal rate occurs with $N_s\sim M$ cosmic strings of tension $\mu_s\sim 1/M$, so an enhancement by a factor $\calo(M)$ over the usual Hawking rate. If corresponding information were  emitted in modes with $\omega\sim 1/R$, this gives an energy density $\sim 1/R^3$ in the immediate vicinity of the horizon -- for a solar mass black hole, around one Planck energy per ${\rm km}^3$.

The real trouble is that if a state with such additional excitations above the Hawking vacuum is traced backwards to its origin, and has evolved via LQFT, it originates with a singular stress tensor at the horizon.  This has come to be known as the ``firewall" problem\AMPS, although the basic issue had been appreciated previously (see, {\it e.g.}, \refs{\SBGmodels, \GiddingsUE}).  To study this, let us introduce lightlike coordinates $x^\pm = t\pm r^*$.  Eq.~\rmdef\ shows that the stress tensor for the outgoing mode  approaches a constant magnitude near the horizon:  specifically, consider $T_{--}$.  However, good coordinates for describing observations of infalling observers crossing the horizon are the Kruskal coordinates, which will be defined as 
\eqn\Krusk{X^\pm = \pm 2R  e^{ \pm{x^\pm/ 2R}}\ .}  
 Two powers of $\partial x^-/\partial X^- = -2R/X^-$ are needed to convert $T_{--}$ to these coordinates; the result diverges at the horizon, which is $x^-=\infty$, or $X^-=0$. 

Thus, if  the necessary information is present in the outgoing modes, without a singular horizon (``firewall"), we need to consider a modification of LQFT evolution, outside the horizon, as proposed in \refs{ \SBGmodels,\GiddingsUE,\NVNL}.  We now turn to a candidate model of such modification, via \lincoup.  

Again working in the interaction picture, with perturbation \lincoup, the Hawking vacuum is modified to
\eqn\Jstate{|J\rangle = T \exp\left\{ -i \int dV_4 J(x)\phi(x)\right\} |0\rangle\ .}
Consider the effect of a source with definite angular momentum, and general time and radial dependence,
\eqn\Jdef{J(x)=  Y_{lm} j_ l(t,r)\ .}
For such a source, and working in terms of the wavepacket basis, \lincoup\ becomes 
\eqn\modecoup{\int dV_4 J \phi = \sum_{j,n} \left[ \ara^\dagger_{jnlm} \int dt dr^* r f(r) j_l(t,r) \ura_{jnl}^*(t,r) 
\right] 
+ \{\ala^\dagger_{jnlm}, \ara_{jnlm}, \ala_{jnlm}\ {\rm terms}\}\ .}
If the source $j_l(t,r)$ has a definite frequency dependence, it will excite modes at the corresponding frequency.  One can also take the source to create excitations at an approximately definite time, via a wavepacket construction such as \wavepdef.  

While a fundamental description of the proposed nonlocal physics may be needed to specify the source's radial dependence, consider for example a radial dependence that A) is smooth at $r=R$, and B) vanishes rapidly at large $r$, e.g. outside the ``atmosphere" at $R<r\roughly< 2R$.  

The first thing to notice is that condition A) means that $ j_l(t,r)$ does not source modes at $r^*\rightarrow-\infty$, since the integral over $r^*$ in this region is rapidly oscillating, leading to cancellation.  One could also introduce a dependence $ j_l(t,r)\sim \exp\{-i\omega(t-r^*)\} \Theta(r-\rho-R)$ (or, corresponding wavepacket), cut off at $r<R+\rho$, with the same result.
 Thus, $T_{--}$ for the state $|J\rangle$ vanishes near the horizon, avoiding a horizon singularity.  There is a $T_{++}$ at the horizon, due to excitation of the ingoing waves of \rmdef, \lmdef, but this does not yield singular behavior for infalling observers.  
$T_{+-}$ is likewise localized away from the horizon.

Clearly $j_l(t,r)$'s satisfying conditions A) and B) can be constructed to excite quite general superpositions of the states $|\{\nra_A\},\{\nla_A\}\rangle$, and in particular clearly sufficiently general to encode the information transfer from the black hole necessary for ultimate unitarity of BH evolution.  We have just seen that such states do not necessarily present violent consequences to infalling observers, if not excited at the horizon.\foot{Such evolution may appear to  violate local energy conservation, though by amounts as small as $1/R$, and without necessarily violating total energy conservation.  A rough analogy is to imagine some new kind of  ``dark matter," whose propagation is governed by a different metric, so which can escape the BH, and which then decays outside the BH into visible matter.}

While necessary information transfer from an isolated black hole can apparently be accommodated by the $l=0$ modes, it is interesting to consider the effects of $j_l(t,r)$'s sourcing higher-l modes.  First, note that for such a mode, the effective potential \effpot\ has a maximum of size $\sim l(l+1)/R^2$, so modes with $\omega^2$ far below this value will have very small magnitude for the gray body factor $B_{l\omega }$.  Consider a source with such a frequency.  Then, the effective (1+1) dynamics has a turning point at $r_> \approx \sqrt{l(l+1)/\omega^2} \gg R$.  A source satisfying condition B) will then have a very small amplitude to excite modes escaping to infinity\foot{Indeed, for a $j_l$ that does not grow rapidly with $r$, the integrand in \modecoup\ also falls rapidly for $r$ above the inner turning point at $r_<$, where the behavior of $\ura_{l,\omega}$ can be estimated by WKB.  Thus, the integral is dominated by the neighborhood of the turning point.  For $l\gg \omega R$, one finds $r_<\approx R + R {(\omega R)^2/ [l(l+1)]}$\ .} -- they will be suppressed by the factor  $\sim B_{l\omega }$.  So, unless the amplitude of the corresponding source $j_{l}$ is adjusted to a large value, such modes will not be appreciably excited -- $l=0$ modes dominate.  Such sources could be present, without large effect.

But now consider a scenario where additional degrees of freedom are introduced, in order to mine the black hole, as discussed in \refs{\UnWa\LaMa-\Frol,\AMPS,\Brown}.  In \Brown, Brown argues that the mining rate cannot exceed the evaporation rate that would occur due to extra modes arising when a collection of cosmic strings thread a black hole.  So, specifically consider the situation with a single cosmic string threading the black hole.  In the field theory approximation, this string is built out of field-theory modes,  so a universal coupling such as \lincoup\ (or, \Tcoup) will couple to the string modes.  The mining enhancement of the black hole evaporation rate arises because the string background introduces new modes with gray body factors of size $\sim 1$.  Then, a coupling of these modes via \lincoup\ (or, \Tcoup), will also produce outgoing excitations, and in contrast to the discussion of the preceding paragraph, these excitations will not be suppressed by the gray body factor.

To summarize the situation, in the presence of general high-l sources, satisfying $\omega\ll l/R$, enhancement of Hawking radiation from introducing a cosmic string is accompanied by the opening of extra channels for information escape, that will be excited by these sources.  This provides a possible mechanism for a black hole to avoid becoming unacceptably ``overfull," namely to have von Neumann entropy $S_{\rm vN}$ exceeding its Bekenstein Hawking entropy $S_{BH}$, in a Gedanken experiment where a cosmic string is introduced to a black hole that is nearly maximally entangled with outgoing radiation.  Rates of information flow will track those of energy flow.  

Of course, another possibility to avoid mining constraints is if information transfer from the black hole interior begins before the Page time\refs{\Page} where 
$S_{\rm vN}(t)$ of Hawking radiation equals $S_{BH}(t)$.  
An early start also needs to happen if information is imprinted into outgoing quanta with $\omega \gg 1/R$; in either case, the actual entropy curve of the outgoing state would be below the envelope given by the minimum of $S_{\rm vN}(t)$ of Hawking radiation and $S_{BH}(t)$.  While  such early initiation of information transfer via couplings \effsource\ offers one possible escape from the potential for overfull black holes\NVNL, due to the possibility of  mining, the present discussion opens a seemingly more natural way to enhance information flow in the presence of mining, and lends support to the possible consistency of the proposal of nonviolent nonlocality.  

Thus, by examining the effect of couplings to modes in the atmosphere of a black hole, in an effective field theory description, we have found that information can transfer to these modes without a singular stress tensor at the horizon, in contrast to the discussion of \AMPS.  Moreover, with modest assumptions about couplings to higher-$l$ modes, we find a mechanism to increase the flux of information-bearing modes when a string is introduced to mine the black hole.  Work is in progress\refs\GiShWIP\  to provide further details, as well as to investigate more general couplings such as \Tcoup.

This paper has not explored details of internal dynamics responsible for the effective sources \effsource\ in the black hole atmosphere.  In particular, there is some latitude in the present parameterization; while a natural frequency is $\omega\sim 1/R$, higher frequency sources have not been shown to be inconsistent.  One may likewise anticipate presence of sources up to a maximum ${\bar l}$ of $l$, {\it e.g.} to account for the possibility of mining.\foot{To meet the maximal mining rate, ${\bar l} \roughly> \sqrt M$.}  An ultimately central question is what dynamics -- beyond LQFT -- describes such ``glistening" of quantum information in the atmosphere of a black hole.  Note also that while there are toy qubit models with equal energy flux to that of Hawking radiation\SBGmodels, the generic model of such behavior -- through \effsource\ -- predicts extra energy flux from a black hole, beyond that of Hawking.

\bigskip\bigskip\centerline{{\bf Acknowledgments}}\nobreak

I thank J. Hartle, D. Marolf, D. Page for helpful conversations, and Y. Shi for helpful conversations and for pointing out a correction to \Tvev.  This work  was supported in part by the Department of Energy under Contract DE-FG02-91ER40618 and by a Simons Foundation Fellowship, 229624, to Steven Giddings.

\listrefs
\end